\documentclass[twocolumn,amsmath,amssymb,appendix,prd,nofootinbib]{revtex4}

\usepackage{graphicx}
\usepackage{hyperref}
\usepackage{dcolumn}
\usepackage{bm}
\usepackage{amsmath}
\usepackage{multirow}

\def\be{\begin{equation}}
\def\ee{\end{equation}}

\def\be{\begin{eqnarray}}
\def\ee{\end{eqnarray}}

\def\bg{\bar{g}}
\def\beq{\begin{eqnarray}}\def\eeq{\end{eqnarray}}
\def\ba#1\ea{\begin{align}#1\end{align}}
\def\bg#1\eg{\begin{gather}#1\end{gather}}
\def\bm#1\em{\begin{multline}#1\end{multline}}
\def\bmd#1\emd{\begin{multlined}#1\end{multlined}}

\def\({\left(}
\def\){\right)}
\def\[{\left[}
\def\]{\right]}

\renewcommand{\t}{\tilde}

\newcommand{\bea}{\begin{eqnarray}}
\newcommand{\eea}{\end{eqnarray}}
\newcommand{\een}{\end{enumerate}}
\newcommand{\bi}{\begin{itemize}}
\newcommand{\ei}{\end{itemize}}
\newcommand{\bc}{\begin{center}}
\newcommand{\ec}{\end{center}}
\newcommand{\bfig}{\begin{figure}}
\newcommand{\efig}{\end{figure}}
\newcommand{\bt}{\begin{table}}
\newcommand{\et}{\end{table}}
\newcommand{\btab}{\begin{tabular}}
\newcommand{\etab}{\end{tabular}}
\newcommand{\bs}{\begin{slide}}
\newcommand{\es}{\end{slide}}

\begin{document}

\preprint{XXX}

\title{Consistent Cosmic Bubble Embeddings}

\author{S.\ Shajidul Haque}
\email {sheikh.haque@uct.ac.za}
\affiliation{Laboratory for Quantum Gravity \& Strings, 
Department of Mathematics \& Applied Mathematics,
University of Cape Town, South Africa}
\author{Bret Underwood}
\email{bret.underwood@plu.edu}
\affiliation{Department of Physics, Pacific Lutheran University, Tacoma, WA 98447, USA} 

\date{\today}

\begin{abstract}
The Raychaudhuri equation for null rays is a powerful tool for finding consistent embeddings of
cosmological bubbles into a background spacetime in a way that is largely independent
of the matter content.
We find that spatially flat or positively curved thin wall bubbles surrounded by a cosmological background
must have a Hubble expansion that is either contracting or expanding slower than the background,
presenting an obstacle for models of local inflation or false vacuum bubble embeddings.
Similarly, a cosmological bubble surrounded by Schwarzschild space must either be contracting (for spatially flat 
and positively curved bubbles)
or bounded in size by the apparent horizon, presenting
an obstacle for embedding cosmic bubbles into a background Universe.
\end{abstract}

\maketitle


\section {Introduction}

Cosmic bubbles -- spherically symmetric homogeneous and isotropic spacetimes surrounded by a spherically symmetric 
background spacetime -- can arise in many interesting applications of General Relativity, 
including local inflation, eternal inflation, cosmological phase transitions, and gravitational collapse.
In these scenarios, spacetime is separated into two separate manifolds ${\mathcal M}^+$ and ${\mathcal M}^-$,
with their own distinct metrics, which are typically joined across a ``thin wall'' hypersurface\footnote{The
assumption of an infinitesimally thin bubble wall does not apply to phase transition bubbles formed in new
inflation \cite{Linde:1981mu}.} using the Israel junction conditions \cite{Israel:1966rt}.

The dynamics of such bubbles can be extremely complicated, depending on the matter content of the 
background and interior of the bubble as well as the tension on the bubble wall and how it interacts
with the surrounding matter.
The most well-understood solutions involve simple assumptions, such as the dust-collapse model of 
Oppenhemier-Snyder \cite{Oppenheimer:1939ue} or false vacuum de-Sitter bubbles in empty space 
\cite{Blau:1986cw,FARHI1990417}.
Some results have also been obtained for interior and exterior spacetimes described by homogeneous and isotropic
FLRW spaces \cite{LagunaCastillo:1986je,Berezin:1987bc,Sakai:1993fu}.
See also
\cite{Berezin:1982ur,Aurilia:1989sb,Adler:2005vn,Aguirre:2005xs,Johnson:2011wt,Wainwright:2013lea,Garriga:2015fdk,Deng:2016vzb} 
for other investigations of cosmic bubble dynamics.

More recently, there has been renewed interest in the onset of local inflation 
arising from inhomogeneous
initial conditions, in which inflation starts in a small patch surrounded by a non-inflating background.
Earlier analytical arguments and numerical work suggested that the inflationary patch will not
grow unless the initial size of the homogeneous patch is larger than the Hubble length
(see e.g.~\cite{Goldwirth:1990iq,Goldwirth:1991rj,Calzetta:1992gv,Perez:2012pn}
and the recent review \cite{Brandenberger:2016uzh}).
These results have been revisited by recent work \cite{East:2015ggf,Clough:2016ymm}, which use
modern numerical relativity codes to study the conditions under which local inflation begins.
While these numerical analyses are not necessarily spherically symmetric, inflation will act to 
homogenize and isotropize the spacetime inside the bubble, so we can view
these models as cosmic bubbles embedded into a larger spacetime.

The dynamics of cosmic bubbles appear to be strongly model-dependent, so it has been difficult to make
general statements about their behavior.
In this paper, we will consider cosmic bubbles from a different angle, by using the
null Raychaudhuri equation to study the consistency of cosmic bubble embeddings.
While we will not be able to derive
dynamical equations of motion for the bubble wall -- which arise from the Israel boundary
conditions -- we will nonetheless find interesting constraints from the null Raychaudhuri equation
on the types of bubble embeddings which are consistent with the propagation of null rays across the bubble wall.
The Raychaudhuri equation 
is particularly powerful because it is independent of specific solutions to the Einstein equations,
and has been used before \cite{Vachaspati:1998dy,Berera:2000xz}
to study local inflation.
In this paper, we will consider more general bubble and background spacetimes beyond inflation.

In particular, we will be interested in the behavior of the expansion $\theta$
of radial, inwardly directed, future-oriented null rays $N^\alpha$. 
For an affinely-parameterized null tangent vector $N^\alpha \nabla_\alpha N^\beta = 0$, 
the expansion $\theta = \nabla_\alpha N^\alpha$ satisfies the null Raychaudhuri equation:
\be
\frac{d\theta}{d\lambda} = -\frac{1}{2} \theta^2 - |\sigma|^2 - R_{\alpha \beta} N^\alpha N^\beta\, ,
\label{Raychaud_1}
\ee
where $\lambda$ is an affine parameter along the geodesic and $\sigma$ is the shear tensor.
The Raychaudhuri equation (\ref{Raychaud_1}) arises from the geometric properties of null vectors,
and as such is independent of the Einstein equations and their solutions.
Imposing the Einstein equations on the last term of (\ref{Raychaud_1}), we have:
\be
\frac{d\theta}{d\lambda} = -\frac{1}{2} \theta^2 - |\sigma|^2 - T_{\alpha \beta} N^\alpha N^\beta\, .
\label{Raychaud_Affine}
\ee
If we assume that the matter inside and outside of the bubble as well as on the bubble wall
obeys the Null Energy Condition (NEC)
$T_{\alpha \beta} N^\alpha N^\beta \geq 0$, the Raychaudhuri
equation (\ref{Raychaud_Affine}) implies that the expansion must be non-increasing,
\be
\frac{d\theta}{d\lambda} \leq 0 \, .
\label{RayNoGo}
\ee

\begin{figure}[t]
\centering\includegraphics[width=.3\textwidth]{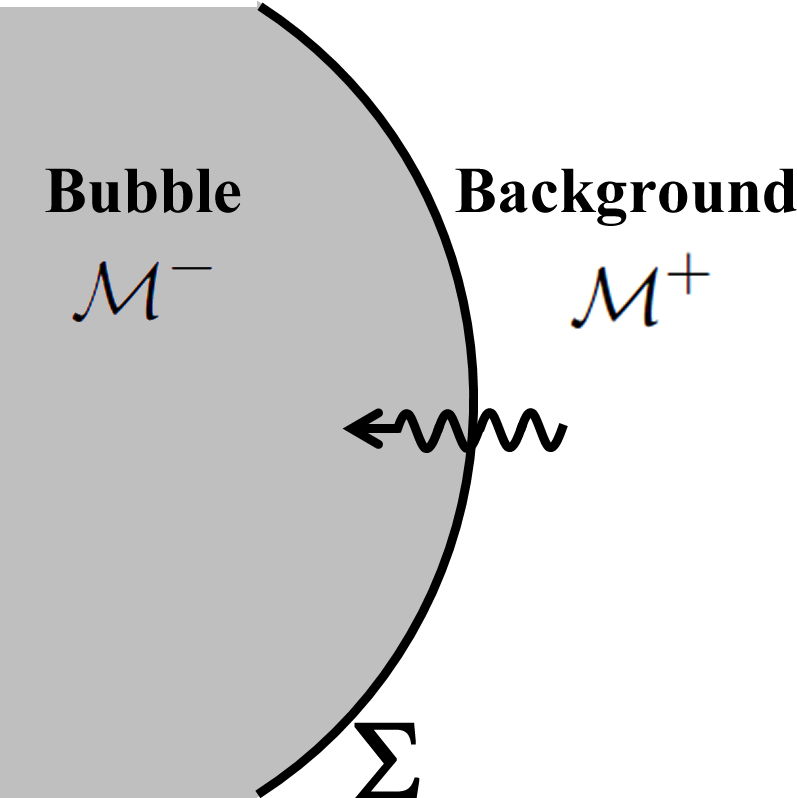}
\caption{We will be considering radially ingoing null rays as they cross the boundary $\Sigma$ from 
a spherically symmetric background into an embedded cosmological bubble.}
\label{fig:ThinWallCrossing}
\end{figure}

As a null ray traverses the bubble wall boundary $\Sigma$ from the background spacetime
into the bubble, as shown in Figure \ref{fig:ThinWallCrossing}, the value of $\theta$ will, in principle, change.
In order for the bubble embedding to be consistent with the Raychaudhuri equation (\ref{RayNoGo}),
the value of $\theta$ must not increase across the wall, 
\be
\Delta \theta = \theta^- - \theta^+\leq 0\, .
\label{thetajump}
\ee
It is common to study cosmic bubbles in the ``thin wall'' limit in which the boundary is infinitesimally thin,
for which (\ref{thetajump}) must be true when evaluated at the (singular) boundary; however, (\ref{thetajump}) 
must also be satisfied for ``thick wall'' bubbles as well, as a null ray leaves the background and enters the bubble.

The Raychaudhuri equation (\ref{Raychaud_Affine}) is independent of specific solutions to the Einstein equations, 
thus (\ref{thetajump}) must be satisfied for any inwardly directed radial null ray, 
independent of the details of the matter content and spacetime structure
of the background and bubble spacetimes as well as the bubble wall, as long such matter obeys the NEC.
As such, our results will be generally applicable to a wide class of bubble embeddings.
We will now consider several different scenarios for the background and bubble spacetimes and 
use (\ref{thetajump}) to constrain the allowed embeddings.


\section{A Bubble in a Cosmological Background}
\label{sec:FRWBackground}

We will begin by considering a spatially flat homogeneous and isotropic cosmological bubble embedded
at the origin of
a homogeneous and isotropic cosmological background; non-spatially flat
bubble and background spacetimes will be considered in Section \ref{sec:NonFlatFRW}. 
Both spacetimes are described by the FLRW metric:
\be
ds_{\pm}^2 = -dt_{\pm}^2 + a_{\pm}(t_{\pm})^2 dr_{\pm}^2 + a_{\pm}(t_{\pm})^2 r_{\pm}^2 d\Omega^2 \, ,
\label{FRWMetricComoving}
\ee
where $\pm$ refers to the background/bubble spacetimes, respectively.

The (timelike) boundary upon which these spacetime regions are joined is the wall of the bubble $\Sigma$.
The metric must be continuous across the bubble wall\footnote{See \cite{Berezin:1987bc,Sakai:1993fu} for an
analysis of the Israel junction conditions across the boundary.}, which implies that the coefficient of $d\Omega^2$
in the metric must be continuous across the boundary $a_+(t_+) r_+|_\Sigma = a_-(t_-) r_-|_\Sigma \equiv R(\tau)$, where  
$\tau$ is the proper time of the wall.
Since the bubble and background spacetimes have (in principle) different cosmological evolutions, 
the comoving radial coordinates $r_{\pm}$ cannot be continuous across the wall. 
The ``areal radius'' $\t r \equiv a(t) r$, however, will be continuous across the wall
and we will set $\t r_+|_\Sigma = \t r_-|_\Sigma = R$ on the bubble wall.

Radial, inwardly directed, future-oriented null rays in the cosmological spacetimes (\ref{FRWMetricComoving})
are described by the (affine) null tangent vector:
\be
N^\alpha_{\pm} = \left( a_{\pm}^{-1}, -a^{-2}_{\pm},0,0\right)
\ee
in either spacetime. The expansion of these rays $\theta^{\pm} = \nabla_\alpha N^\alpha_{\pm}$ is then
\be
\theta^{\pm} = \frac{2}{a_{\pm}} \left(H_{\pm} - \frac{1}{\tilde r}\right)\, .
\label{thetaAffineFRW}
\ee
Following \cite{Hayward:1993wb,Faraoni:2011hf}, (\ref{thetaAffineFRW}) defines an 
\emph{apparent horizon}
for the bubble and background spacetimes where $\theta^{\pm} = 0$:
\be
\tilde r_{AH}^{\pm} = H_{\pm}^{-1}\, .
\ee
Null rays that are at an areal distance from the origin that is smaller than the radius of the apparent horizon 
$\tilde r < \tilde r_{AH}^{\pm}$ have negative expansion, as is expected for converging rays.
However, null rays that are at a areal distance larger than the radius of the apparent horizon $\tilde r > \tilde r_{AH}^{\pm}$
have a positive divergence, indicating that the expansion of space is overcoming the expected geometric convergence of the rays.

\subsection{Crossing the Bubble Wall} 
\label{sec:WallCrossFRWFlat}

In the limit of an infinitesimally thin bubble wall, as a null ray crosses the boundary the radial coordinate is continuous 
$\t r_{\pm}|_\Sigma = R$ but the scale factor $a_{\pm}$ and Hubble expansion rate $H_{\pm}$ are not.
Thus, the requirement that $\theta$ must decrease (\ref{thetajump}) becomes:
\be
\Delta \theta =  \frac{2}{a_-}\left(H_- - \frac{1}{R}\right) -\frac{2}{a_+}\left(H_+ - \frac{1}{R}\right) \leq 0\, .
\label{thetajumpAffineFRW}
\ee
The metrics (\ref{FRWMetricComoving}) contain an ambiguity: it is always possible to make a simultaneous rescaling of the scale
factor and comoving radial coordinate by a constant $a_{\pm} \rightarrow \lambda_{\pm} a_{\pm}, r_{\pm} \rightarrow \lambda_{\pm}^{-1} r_{\pm}$ 
that leaves the metric and physical distances invariant.
We will use this freedom to fix the bubble and background scale factors to be equal to one 
at the time of the null ray wall crossing $t_{\pm,cross}$ for
the respective spacetimes only, e.g.~
$a_+(t_{+,cross}) = a_-(t_{-,cross}) = 1$. 
Since the matter content of the background and bubble generically differ, this implies that the scale factors should
typically not be the same at any other time. Utilizing this freedom, we can simplify (\ref{thetajumpAffineFRW}) to:
\be
\Delta \theta = \theta^- - \theta^+ = 2(H_-  -H_+) \leq 0\, .
\label{thetaFAILFRW}
\ee
This constraint has important implications for the embedding of cosmological bubbles inside cosmological background spacetimes.

If the background is expanding but the bubble spacetime is collapsing $H_- < 0$, then we are able to satisfy
(\ref{thetaFAILFRW}) without any difficulty, indicating that a collapsing bubble inside
an expanding background is always an allowed solution.
Alternatively, if the background spacetime is collapsing, then (\ref{thetaFAILFRW}) indicates that the bubble spacetime cannot
be expanding while still satisfying the Raychaudhuri equation. Since this is of little practical cosmological interest for us, however,
we will not consider this situation further.

If both the bubble and the background spacetimes are expanding $H_{\pm} > 0$, then (\ref{thetaFAILFRW}) is only satisfied
for a bubble that has a lower Hubble rate than the background $H_- < H_+$, indicating
that the energy density of the bubble is smaller than the background $\rho_- < \rho_+$. Some simple examples
include a
true vacuum bubble embedded inside a false vacuum background, or a low temperature bubble inside a higher temperature
background.
However, (\ref{thetaFAILFRW}) fails for a bubble that has a higher Hubble rate than the
background $H_- > H_+$, including false vacuum or high temperature bubbles.
This represents a significant challenge for embedding false vacuum or ``hot'' bubbles into cosmological backgrounds.
This further extends the results of \cite{Berezin:1987bc,Sakai:1993fu}, which found that
the Israel junction conditions do not allow for a spatially flat bubble with $H_- > H_+$ unless
the bubble is superhorizon-sized.
We find a stronger result here, which is that a spatially flat bubble with $H_- > H_+$ is not allowed for
any size bubble.
We emphasize that this result is independent of the details of the matter content inside and outside of the
bubble as well as the tension of the bubble wall, as long as the matter obeys the NEC.
The Raychaudhuri equation thus provides a very strong constraint on allowed bubble embeddings.

A limited form of this result was found in \cite{Vachaspati:1998dy}, which considered
an inflationary bubble embedded inside a non-inflationary cosmological background with $H_- > H_+$.
\cite{Vachaspati:1998dy} noticed that for inflationary bubbles that are larger than their own apparent horizon
but smaller than the apparent horizon of the background $\tilde r_{AH}^- < R < \tilde r_{AH}^+$,
an ingoing light ray traversing the bubble wall that starts in the background will begin with 
negative expansion $\theta^+ < 0$.
However, after traversing the bubble wall boundary, the expansion is now positive $\theta^- > 0$ and therefore 
$\theta$ is non-decreasing, in violation of the Raychaudhuri equation.
This change in sign of $\theta$ can be avoided if the inflationary bubble is larger
than both the bubble and background apparent horizons $R > \tilde r_{AH}^+,\tilde r_{AH}^-$, so that
$\theta^{\pm}$ is positive in both spacetimes \cite{Vachaspati:1998dy}.
Putting aside the difficulty of establishing such a configuration in a causal way, our result (\ref{thetaFAILFRW}) indicates 
that it is not sufficient for
$\theta^{\pm}$ to simply be positive in both spacetimes, as $\theta^-$ for the inflationary
bubble is still larger than $\theta^+$ for the background when $H_- > H_+$ for any size bubble.

The challenge presented by (\ref{thetaFAILFRW}) to bubbles with $H_- > H_+$ is quite basic, and
it is compelling to view the failure of (\ref{thetaFAILFRW}) as due to the unrealistic assumption 
that the bubble and background cosmological
spacetimes (\ref{FRWMetricComoving}) are glued together on an infinitesimally thin wall.
Since the gravitating energy enclosed by the bubble 
is larger than it would be if filled with the background energy density, we expect the
energy density of the bubble to locally
backreact on the background metric so that it deviates from the pure cosmological form.
Indeed, for a pure de-Sitter bubble embedded in a background that is vacuum \cite{FARHI1990417} 
or dominated by a positive cosmological constant  \cite{Aguirre:2005xs},
the spacetime is approximately Schwarzschild in the vicinity of the bubble wall.
Unfortunately, analytic forms for the backreaction of the bubble energy density are not known
when the bubble and background are filled with a generic cosmological fluid.

\begin{figure}[t]
\centering\includegraphics[width=.3\textwidth]{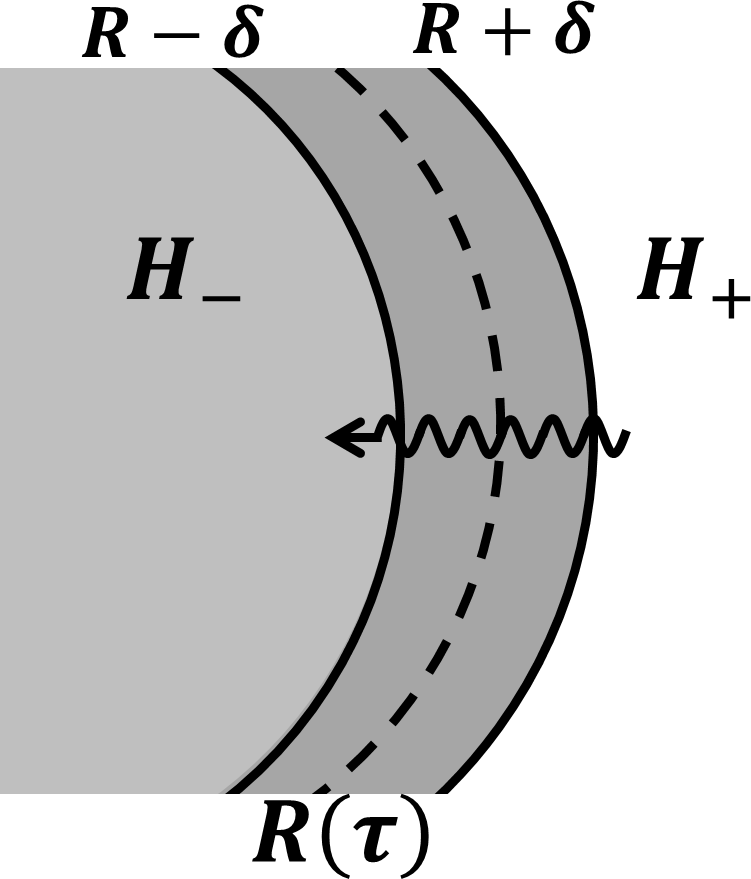}
\caption{A null ray traveling across a ``thick wall'' boundary between a cosmological background $H_+$ and bubble $H_-$ 
leaves the background at $\t r = R+\delta$ and enters the bubble at $\t r = R-\delta$.}
\label{fig:ThickWallCrossing}
\end{figure}

Nevertheless, we can move beyond our approximation of an infinitesimally thin wall in a generic way by assuming that the 
bubble wall now has a ``thickness'' of $2\delta$, as in Figure \ref{fig:ThickWallCrossing}.
We will leave the spacetime geometry inside the thick wall unspecified, as its form likely will include the unknown backreaction of the bubble and tension of the wall.
Since we do not know the details of the metric inside of the thick wall, we cannot compute $\theta$ inside the wall. 
Nevertheless, it must still be true that (\ref{thetajump}) is satisfied for a null ray as it enters and exits the thick wall.

An ingoing null ray enters the thick wall from the background at $\tilde r = R + \delta$ and leaves the wall into the bubble at $\tilde r = R-\delta$,
so we have\footnote{We have again used the rescaling freedom in the scale factor to set $a_+ = a_- = 1$
when the ray crosses the respective exterior and interior boundaries.}:
\be
\Delta \theta = 2 (H_--H_+) -\frac{4 \delta} {R^2-\delta^2} \leq 0\, .
\label{thetaThick}
\ee
In contrast to (\ref{thetaFAILFRW}), it is now possible to solve (\ref{thetaThick}) for an expanding bubble with $H_- > H_+$.

In particular, (\ref{thetaThick}) is satisfied if the thickness of the wall is larger than
\be
\delta \geq \frac{1}{2} R^2 \Delta H\, ,
\label{thickBound}
\ee
where $\Delta H = H_- - H_+$, and we assumed\footnote{
If $R \Delta H \sim {\mathcal O}(1)$, then the
wall must be approximately the same size as the bubble itself $\delta \geq \Delta H \sim R$ in order to solve
(\ref{thetaThick}), which implies
that the bubble spacetime is almost certainly not described by a homogeneous and isotropic cosmological metric.}
$R |\Delta H| \ll 1$.

It is interesting to see how the presence of a ``thick'' wall satisfying (\ref{thickBound}) also evades the argument of \cite{Vachaspati:1998dy}
described above. In the presence of a thick wall of size $\t r_{AH}^- < R < \t r_{AH}^+$ satisfying (\ref{thickBound}),
a null ray exits the wall and enters the bubble cosmology at the inner boundary of the wall $\tilde r = R- \delta$.
Since 
$H_- (R-\delta) \leq H_-R \left(1-\frac{1}{2} H_- R \left(1-\frac{H_+}{H_-}\right)\right) \leq 1-\frac{1}{2}\left(1-\frac{H_+}{H_-}\right) < 1$,
the inner boundary of the wall is smaller than the bubble apparent horizon $R-\delta < H_-^{-1} = \tilde r_{AH}^-$.
Thus, the expansion of the null ray is negative when it enters the cosmological part of the bubble, and Raychaudhuri's equation
(\ref{thetaThick}) can be satisfied.

It is important to note that the condition (\ref{thickBound}) is a necessary condition for the thickness of a wall surrounding
a cosmological bubble with $H_- > H_+$, but it is not sufficient.
Our approach has avoided specifying the behavior of $\theta$ inside the ``thick'' bubble wall, 
and it must be the case that $\theta$
is decreasing inside the thick bubble wall in a way that interpolates between the expansions $\theta$
of the exterior and interior spacetimes to satisfy the Raychaudhuri equation.
In Section \ref{sec:Schwarz} we consider a cosmological bubble surrounded by a Schwarzschild spacetime,
which can serve as one possible model for the spacetime inside the ``thick wall.''
In the next subsection, we generalize our argument to include non-zero spatial curvature for
both the bubble and the background, finding that the main results of this section hold
for a larger range of sings of the spatial curvature of the bubble and background.

\subsection{Non-zero Spatial Curvature for Bubble and Background}
\label{sec:NonFlatFRW}

The constraint (\ref{thetaFAILFRW}), while independent of the matter content of the bubble and background,
applies only to bubble and background spacetimes that are spatially flat.
In general, however, the bubble and background can have their own distinct spatial curvature.
Indeed, \cite{Berera:2000xz} generalized the argument of \cite{Vachaspati:1998dy} to non-flat spatial sections and found that 
it is possible to avoid changes of sign of $\theta^\pm$ when crossing the boundary if the bubble is initially negatively curved.
In this section, we will generalize our results from Section \ref{sec:WallCrossFRWFlat} 
to non-flat spatial geometries by requiring that $\theta$ be non-increasing.

Including spatial curvature in a homogeneous and isotropic spacetime, we start with the metrics:
\be 
ds_{\pm}^2=-dt_{\pm}^2 +a_{\pm}(t)^2 \left( \frac{dr_{\pm}^2}{1-k_{\pm} r_{\pm}^2} + r_{\pm}^2 d\Omega^2 \right)\, ,
\label{eq:NonFlatFRW}
\ee
where $k_{\pm} > 0$ ($k_{\pm} < 0$) corresponds to positive (negative) spatial curvature, and $k_{\pm} = 0$ is flat
space. Note that (\ref{eq:NonFlatFRW}) still allows for an independent constant rescaling of 
the scale factors $a_{\pm} \rightarrow \lambda_{\pm} a_{\pm}$ as long as the comoving radial coordinates and spatial
curvatures are rescaled as well $r_{\pm} \rightarrow \lambda_{\pm}^{-1} r_\pm, k_{\pm} \rightarrow \lambda_{\pm}^2 k_{\pm}$.
This is only consistent if one does not choose $k_{\pm} = \pm 1$, as is common in the presence 
of spatial curvature, which we will avoid.

Radially inward affine null rays in the spacetimes (\ref{eq:NonFlatFRW}) take the form:
\be
N_{\pm}^\alpha = \frac{1}{a_{\pm}}\left(1,-\frac{\sqrt{1-k_{\pm} r_{\pm}^2}}{a_{\pm}},0,0\right)\, ,
\label{eq:NonFlatNull}
\ee
with corresponding expansion $\theta^\pm = \nabla_\alpha N_\pm^\alpha$,
\be
\theta^{\pm} = \frac{2}{a_{\pm}} \left(H_{\pm} - \frac{\sqrt{1-k_{\pm} r_{\pm}^2}}{\t r}\right)\, ,
\ee
where again we have used the areal radius $\t r = a_- r_- = a_+ r_+$, which is continuous across the boundary,
even in the presence of spatial curvature.
As in flat space, a vanishing expansion for ingoing null rays $\theta^{\pm} = 0$ defines an apparent horizon
for non-spatially flat FLRW space:
\be
\t r^{\pm}_{AH} = \frac{1}{\sqrt{H_\pm^2 + k_\pm/a_\pm^2}}\, .
\label{eq:AHNonFlat}
\ee
For $\t r > r_{AH}^\pm$, the expansion is positive due to the expansion of space.

As the null ray crosses the boundary $\t r|_\Sigma = R$, 
the requirement from the Raychaudhuri equation (\ref{thetajump}) that $\theta$ must be non-increasing
$\Delta \theta = \theta^- - \theta^+ \leq 0$ becomes:
\be
\label{eq:thetajumpNonFlat}
&&\Delta \theta = \\
&& \frac{2}{a_-} \left(H_- - \frac{\sqrt{1-k_- r_-^2}}{R}\right) - \frac{2}{a_+} \left(H_+ - \frac{\sqrt{1-k_+ r_+^2}}{R}\right)  \nonumber \\
&=& 2 (H_- - H_+) - \frac{2}{R} \left(\sqrt{1-k_- r_-^2} - \sqrt{1-k_+ r_+^2}\right) \leq 0\, ,\nonumber 
\ee
where we again used our rescaling freedom to set $a_{\pm} = 1$ at the wall crossing.
It does not seem possible to make general statements about the implications of (\ref{eq:thetajumpNonFlat}),
so we will examine the constraints imposed by (\ref{eq:thetajumpNonFlat}) for specific cases.
The results are summarized in Table \ref{table:NonFlatConstraints}.

In particular, for a background that is either flat or negatively curved $k_+ \leq 0$, we have $\sqrt{1-k_+ r_+^2} \geq 1$,
while a bubble that is either flat or positively curved $k_- \geq 0$ has $\sqrt{1-k_- r_-^2} \leq 1$.
Combining these two inequalities with (\ref{eq:thetajumpNonFlat}), we find
\be
H_- - H_+ \leq 0\, .
\label{eq:NonFlatColdOnly}
\ee
This is an analogous constraint as (\ref{thetaFAILFRW}) 
from Section \ref{sec:WallCrossFRWFlat}: for an expanding bubble and background, only bubbles with a smaller Hubble
rate than that of the background are consistent with the Raychaudhuri equation.
(Similar arguments as given in Section \ref{sec:WallCrossFRWFlat} hold for a non-expanding bubble or background).

For a background with positive spatial curvature $k_+ > 0$ and positive or flat spatial curvature
for the bubble $k_- \geq 0$, we can rearrange (\ref{eq:thetajumpNonFlat}) into the form
\be
H_- - H_+ \leq \frac{1}{R}\left(\sqrt{1-k_- r_-^2} - \sqrt{1-k_+ r_+^2}\right) \leq \frac{1}{R}\, ,
\label{eq:kPlusPosConstraint}
\ee
where the last inequality follows because the square roots are bounded for non-negative $k_{\pm}$.
This constraint (\ref{eq:kPlusPosConstraint}) is considerably less stringent than (\ref{eq:NonFlatColdOnly})
since while an expanding bubble with a Hubble rate smaller than that of the background still satisfies 
(\ref{eq:kPlusPosConstraint}), we can now have an expanding bubble with a Hubble rate {\it larger} than
that of the background, provided that the size of the bubble is smaller than $R \leq (H_- - H_+)^{-1}$.
If the bubble and the background expansion rates are not too different, this does not amount to too
stringent of a constraint. However, if the bubble is expanding much faster than the background $H_- \gg H_+$,
then the bubble size is bounded above by the bubble's inverse Hubble length $R < H_-^{-1}$.
Similarly, if the background itself is collapsing $H_+ < 0$, the size of the bubble is again constrained by
a combination of the expansion rates. Alternatively, if the bubble is collapsing, (\ref{eq:kPlusPosConstraint})
is automatically satisfied, as in the flat space case.

Finally, to enumerate all possibilities we must consider the cases when the bubble is negatively curved $k_- < 0$.
In all of these cases, (\ref{eq:thetajumpNonFlat}) implies a bound on the size of the bubble:
\be
R \leq \frac{\sqrt{1+|k_-|\ r_-^2} - \sqrt{1-k_+\ r_+^2}}{H_- - H_+}\, ,
\label{eq:kMinusNegConstraint}
\ee
(where we took the absolute value of the bubble spatial curvature for clarity).

It is interesting to compare the results of Table \ref{table:NonFlatConstraints} with the constraints
obtained from the Israel boundary conditions on FLRW spacetimes embedded in FLRW backgrounds
from \cite{Berezin:1987bc,Sakai:1993fu}.
In particular, \cite{Sakai:1993fu} finds that there
are no restrictions on the bubble embedding when the difference in energy density between the background and
bubble is positive $\rho^+ > \rho^-$ and larger than the surface energy density ${\mathcal S}$ of the 
bubble wall $\rho^+ - \rho^- > 6\pi G {\mathcal S}$.
However, when this difference is either negative, as when $\rho^- > \rho^+$, or smaller than
the surface energy density, then it is not possible to satisfy the Israel junction conditions for flat or negatively
curved spacetimes $k_+ \leq 0$ (for any value of the bubble spatial curvature) unless the bubble is larger than 
super-Hubble size $R > H_+^{-1}$.
In contrast, we find a much stronger result since if the background is flat or negatively curved and the bubble
is flat or positively curved,
the embedding is inconsistent with the NEC and the
Raychaudhuri equation unless the bubble expansion rate is smaller than the background $H_- < H_+$, for
any sized bubble.
On the other hand, if the bubble is negatively curved, our constraint (\ref{eq:kMinusNegConstraint}) appears 
to put an upper bound on the size of an allowed bubble, rather than the lower bound of \cite{Sakai:1993fu}.
Thus, we see that the constraints imposed by the Raychaudhuri equation are complementary, and in some cases
much stronger, than constraints obtained by the Israel boundary conditions.

\setlength{\tabcolsep}{0.5em} 

\begin{table}[t] 
\centering \begin{tabular}{c| c |c c} 
& \multicolumn{3}{c}{Background} \\
Bubble& $k_+ > 0$ & $k_+ = 0$ & $k_+ < 0$ \\\hline
$k_- > 0$ & \multirow{2}{*}{$H_- - H_+ \leq R^{-1}$} &\multicolumn{2}{|c}{\multirow{2}{*}{$H_- < H_+$}} \\
$k_- = 0$ & \hspace{1.2in} & \multicolumn{2}{|c}{\hspace{1.2in}}\\  \hline 
$k_- < 0$ &  \multicolumn{3}{c}{$R(H_- - H_+) \leq \sqrt{1+|k_-|\ r_-^2} - \sqrt{1-k_+\ r_+^2}$} \\
\end{tabular}
\caption{Requiring that the expansion $\theta$ of a null ray be non-increasing as the ray traverses the boundary
between a background spacetime and bubble spacetime when including spatial curvature leads
to different constraints depending on the relative signs of the spatial curvatures of the spaces.} 
\label{table:NonFlatConstraints} 
\end{table}

As in Section \ref{sec:WallCrossFRWFlat}, it is tempting to see the failure of the Raychaudhuri equation
for non-spatially flat bubbles as due to a possibly unrealistic assumption that the size of the bubble wall is
infinitesimally thin.
We can generalize our argument here to include a wall with a thickness of $2\delta$ as in Figure \ref{fig:ThickWallCrossing}
with unspecified geometry, so that a null ray leaves the background and enters the wall at $\tilde r = R + \delta$
and leaves the wall and enters the bubble at $\tilde r = R- \delta$.
The expansion $\theta$ must be non-increasing from when the null ray leaves the background and enters the bubble wall,
\be
\Delta \theta = \theta^-|_{\t r = R-\delta}- \theta^+|_{\t r = R+\delta}
\ee
leading to the constraint:
\be
\label{eq:ThickWallNonFlat}
&&\frac{2}{a_-(t^*_-)} \left(H_- - \frac{\sqrt{1-k_- (R-\delta)^2/a_-^2}}{R-\delta}\right)  \\
&&	- \frac{2}{a_+(t^*_+)}\left(H_+ - \frac{\sqrt{1-k_+ (R+\delta)^2/a_+^2}}{R+\delta}\right) \leq 0\, ,\nonumber
\ee
where $t^*_{\pm}$ are the times when the null ray crosses the boundary out of (into) the background (bubble) spacetimes.
It is difficult to draw generic conclusions from (\ref{eq:ThickWallNonFlat}). 
Specific constraints can be obtained in the $H_+ \rightarrow 0$ limit, though since in this limit the background is empty,
so we should consider a Schwarzschild background spacetime instead.


\section{Bubble in a Schwarzschild Background}
\label{sec:Schwarz}

\begin{figure}[t]
\centering \includegraphics[width=.45\textwidth]{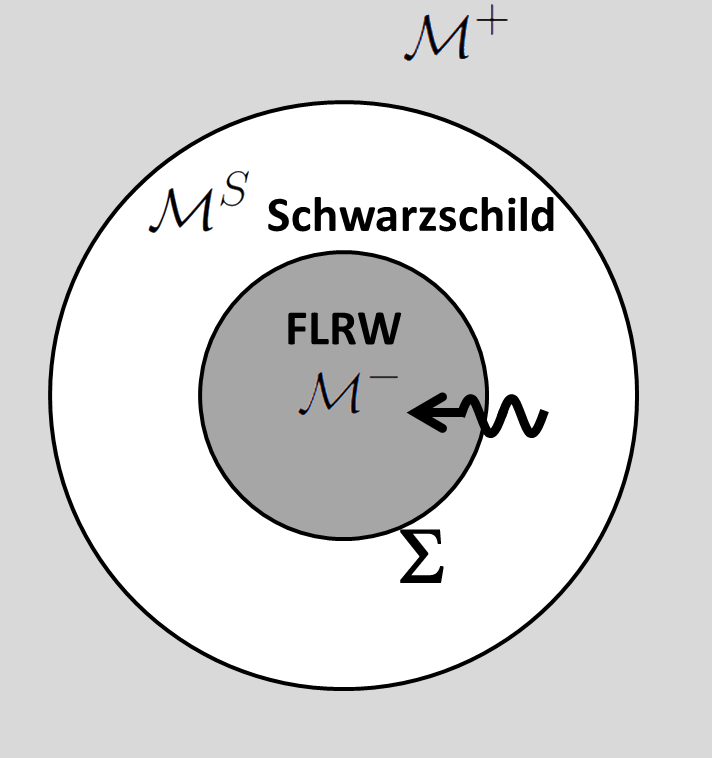}
\caption{A FLRW bubble surrounded by a Schwarzschild background can serve as a model for a ``vacuole''
embedded in a larger expanding background. We study the behavior of radial null rays as they
traverse from the Schwarzschild background into the bubble.}
\label{fig:vacuole}
\end{figure}

We will now consider our cosmological bubble to be surrounded by a background which is vacuum Schwarzschild spacetime.
This can serve as a model of a false vacuum bubble in flat space, as in \cite{Blau:1986cw,Farhi:1986ty,FARHI1990417},
or as a Schwarzschild ``vacuole'' embedded in a larger cosmological background, as in Figure \ref{fig:vacuole}.
Indeed, \cite{Garriga:2015fdk,Deng:2016vzb} have argued that vacuum bubbles will develop
just such a Schwarzschild layer as they evolve in a post-inflationary Universe.
This Schwarzschild layer can thus serve as a simple model for the ``thick wall'' introduced in Section \ref{sec:FRWBackground},
or as a background for the bubble in its own right.

We will thus take the background spacetime to be spatially flat Schwarzschild 
space:
\be
ds_S^2 = && -\left(1-\frac{2GM}{\t r_S}\right)dt_S^2 + \left(1-\frac{2GM}{\t r_S}\right)^{-1} d\t r_S^2 \nonumber \\
&& + \t r_S^2 d\Omega^2\, ,
\ee
where we will use a subscript ``$S$'' for the background to avoid confusion with the FLRW background coordinates from the previous
section.
A radially ingoing (affine) null ray in this spacetime,
\be
N_{S}^\alpha = \left(\left(1-\frac{2GM}{\t r_S}\right)^{-1},-1,0,0\right)\, ,
\ee
has the expansion:
\be
\theta^{S} = \nabla_\alpha N_{S}^\alpha = -\frac{2}{\t r_S}\, ,
\ee
which is what we would expect to get by setting $H_+ = 0$ in the FLRW background of the previous section.

As the null ray traverses the bubble wall, the spacetime changes from the exterior Schwarzschild space to 
the interior FLRW cosmology (we will consider $k_- \neq 0$ here):
\be
ds^2 = -dt_-^2 + \frac{a_-(t_-)^2}{1-k_-r_-^2} dr_-^2 + a_-(t_-)^2 r_-^2 d\Omega^2\, .
\ee
As before, we will find it convenient to work with the areal radius $\t r_- = a_-(t_-) r_-$, which is continuous across
the boundary of the bubble.
A radially ingoing affine null ray for this non-spatially-flat FLRW spacetime takes the form
\be
N_-^\alpha = \frac{1}{a_-} \left(1,-a_-^{-1} \sqrt{1-k_- r_-^2},0,0\right)\, ,
\ee
which has the expansion
\be
\theta^- = \frac{2}{a_-} \left(H_- - \frac{\sqrt{1-k_- \t r_-^2/a_-^2}}{\t r_-}\right)\, .
\ee
In order for the Raychaudhuri equation (\ref{thetajump}) 
to be satisfied as the null ray crosses the thin wall boundary $\t r_S|_\Sigma =\tilde r_-|_\Sigma  = R$, we must have,
\be
\Delta \theta &=& \theta^- - \theta^{S} \nonumber \\
&=& \frac{2}{a_-} H_- + \frac{2}{R}\left(1-\frac{\sqrt{1-k_- R^2/a_-^2}}{a_-}\right) \leq 0\, .
\label{thetaSchw}
\ee

Note that since $\theta^-$ is positive when the size of an expanding bubble is larger than its apparent
horizon $R > \t r_{AH}^-$, while $\theta^S$ is always negative, the requirement from the Raychaudhuri equation
that $\theta$ be non-increasing implies that the bubble may never be larger than its own apparent horizon.
Thus, an arbitrarily large expanding bubble cannot develop when surrounded by a Schwarzschild spacetime.
It is interesting to compare this result to that of \cite{Farhi:1986ty}, which found that an expanding bubble
larger than its apparent horizon must have begun in an initial singularity.
We find a complementary result here, that such a bubble embedding would not be consistent with the Raychaudhuri
equation to begin with.

Utilizing our freedom to rescale the scale factor of the bubble to $a(t_-^*) = 1$ at the time
of crossing $t_-^*$, the condition (\ref{thetaSchw}) becomes
\be
2 H_- + \frac{2}{R}\left(1-\sqrt{1-k_- R^2}\right) \leq 0\, .
\label{thetaSchw1}
\ee
As in the previous section, we can impose stronger constraints from (\ref{thetaSchw1}) for specific assumptions about the
spatial curvature of the bubble. Assuming a bubble that is either spatially flat or positively curved
$k_- \geq 0$, the second term of (\ref{thetaSchw1}) is always positive; thus (\ref{thetaSchw1}) requires
that the bubble spacetime be contracting
\be
H_- \leq 0\, .
\label{SchwBoundFlat}
\ee
This implies that a flat or positively curved cosmic bubble 
embedded inside of a Schwarzschild background must be collapsing in order to satisfy
the Raychaudhuri equation.
This is consistent with the the $k_- \geq 0$ Oppenheimer-Synder solutions \cite{Oppenheimer:1939ue}
for a ball of collapsing dust, some of the first
and most famous solutions of cosmological bubbles embedded in a Schwarzschild spacetime background.
While the Oppenheimer-Synder solution is specifically for pressureless dust and zero surface tension, our
result (\ref{SchwBoundFlat}) does not make any assumptions about the matter content of the bubble or surface tension
(as long as it satisfies the NEC), thus generalizing the conditions under which the bubble collapses.

One motivation for considering a background described by a Schwarzschild spacetime was that it could
serve as a simple model of a ``thick wall'' between a cosmological FLRW background and the bubble.
Because of this, we have assumed the thin wall approximation for the bubble wall.
However, it can be the case that the boundary between the Schwarzschild and bubble spacetimes 
is also itself thick; we will therefore
generalize our results for a thick wall of size $2\delta$, as in Section \ref{sec:FRWBackground}.

First, note that even in the presence of a thick wall, the inner boundary of the wall $R-\delta$ must be smaller than the
apparent horizon of the bubble, for the same reasons as described below (\ref{thetaSchw}), again indicating
that the bubble size may not be larger than its apparent horizon for any value of the bubble spatial curvature.
Constraints on the minimum thickness of a sub-horizon bubble wall 
can be obtained from (\ref{eq:ThickWallNonFlat}) by setting $H_+ = 0$ and $k_+ = 0$
(since $\theta^S$ and $\theta^+$ agree in this liimit).
For a spatially flat bubble, the thickness of the bubble wall must be larger than
\be
\delta > H_-^{-1} \left(-1+\sqrt{1+R^2H_-^2}\right)\, ,
\ee
in order to satisfy the Raychaudhuri equation. For bubbles much smaller than their apparent horizon
size $R H_- \ll 1$, the required thickness $\delta_{min}$ is small compared to the size of the bubble 
$\delta_{min}/R = \frac{1}{2} H_- R \ll 1$. However, for bubbles that are comparable in size
their apparent horizon $H_- R \sim 1$, the wall thickness is comparable to the size
of the bubble itself $\delta \sim R$, necessitating a different spacetime structure for the entire bubble.
Thus, even with a ``thick wall'' with an unspecified metric, large expanding bubbles embedded into Schwarzschild space are
inconsistent with the Raychaudhuri equation.

\section{Discussion}

The Raychaudhuri equation requires that the expansion of radially inward null rays must be non-increasing
as long as matter obeys the Null Energy Condition (NEC), independent
of the Einstein equations.
We have used this to study allowed spherically symmetric embeddings of FLRW cosmological bubbles into
various background spacetimes.

We found that when the background is FLRW space with non-positive spatial curvature $k_+ \leq 0$
and the bubble has non-negative spatial curvature $k_- \geq 0$, 
the Raychaudhuri equation constrains the bubble's Hubble expansion rate to be smaller than that of the 
background's, $H_- < H_+$, for any size bubble.
(For other combinations of the spatial curvature of the bubble and background, see Table \ref{table:NonFlatConstraints}.)
This result has several important implications for cosmological bubbles. 
In particular, it rules out a broad class of embeddings 
of false vacuum bubbles into true vacuum backgrounds obtained by
joining the spacetimes together across the bubble wall, irrespective of the details
of the matter content on the wall as long as the matter obeys the NEC.
It also implies a difficulty with embedding local inflation into a larger cosmological background,
as first suggested by \cite{Vachaspati:1998dy} in a more limited context.
In particular, since the Hubble expansion rate in a near-dS inflating bubble is likely to be larger
than that of its non-dS background (since the latter will decrease with time while the former does not),
a model of local inflation obeying the NEC conflicts with this constraint.
This has important implications for recent studies of the onset of inflation from inhomogeneous
initial conditions \cite{East:2015ggf,Clough:2016ymm}.

We also considered bubbles surrounded by empty Schwarzschild space, which could
serve as a model of a ``vacuole'' embedded in a larger FLRW background.
We found that expanding bubbles must be smaller than their own apparent horizon in order to satisfy
the Raychaudhuri equation. 
This rules out the possibility of embedding an expanding bubble Universe
in flat space, extending the result \cite{Farhi:1986ty} that such a bubble must start from an initial singularity.
Further, we show that a bubble of any size with flat or positives spatial curvature cannot be expanding but rather
must be contracting, generalizing the collapsing behavior of the Oppenheimer-Synder dust ball solution 
\cite{Oppenheimer:1939ue}
to any matter content satisfying the NEC.

Our results place strong limits on the embedding of spherically symmetric bubbles
into a background spacetime, so we also considered the relaxation of the infinitesimal ``thin wall'' approximation
for the bubble wall.
Without specifying the metric within the ``thick wall'' of the bubble, the Raychaudhuri equation requires that the expansion
of the null rays be non-increasing between the outer and inner boundaries of the bubble wall.
For a cosmological background, we derived a lower bound on the required thickness of the bubble
wall to satisfy the Raychaudhuri equation, and showed that for a spatially flat background and bubble the
inner boundary of the bubble is smaller than the apparent horizon of the bubble spacetime, so that again
the bubble may not be larger than its own apparent horizon.
For a Schwarzschild background the bubble must be much smaller than its own apparent horizon or else the required thickness
of the wall is comparable to the size of the bubble itself.

\begin{figure}[t]
\centering\includegraphics[width=.47\textwidth]{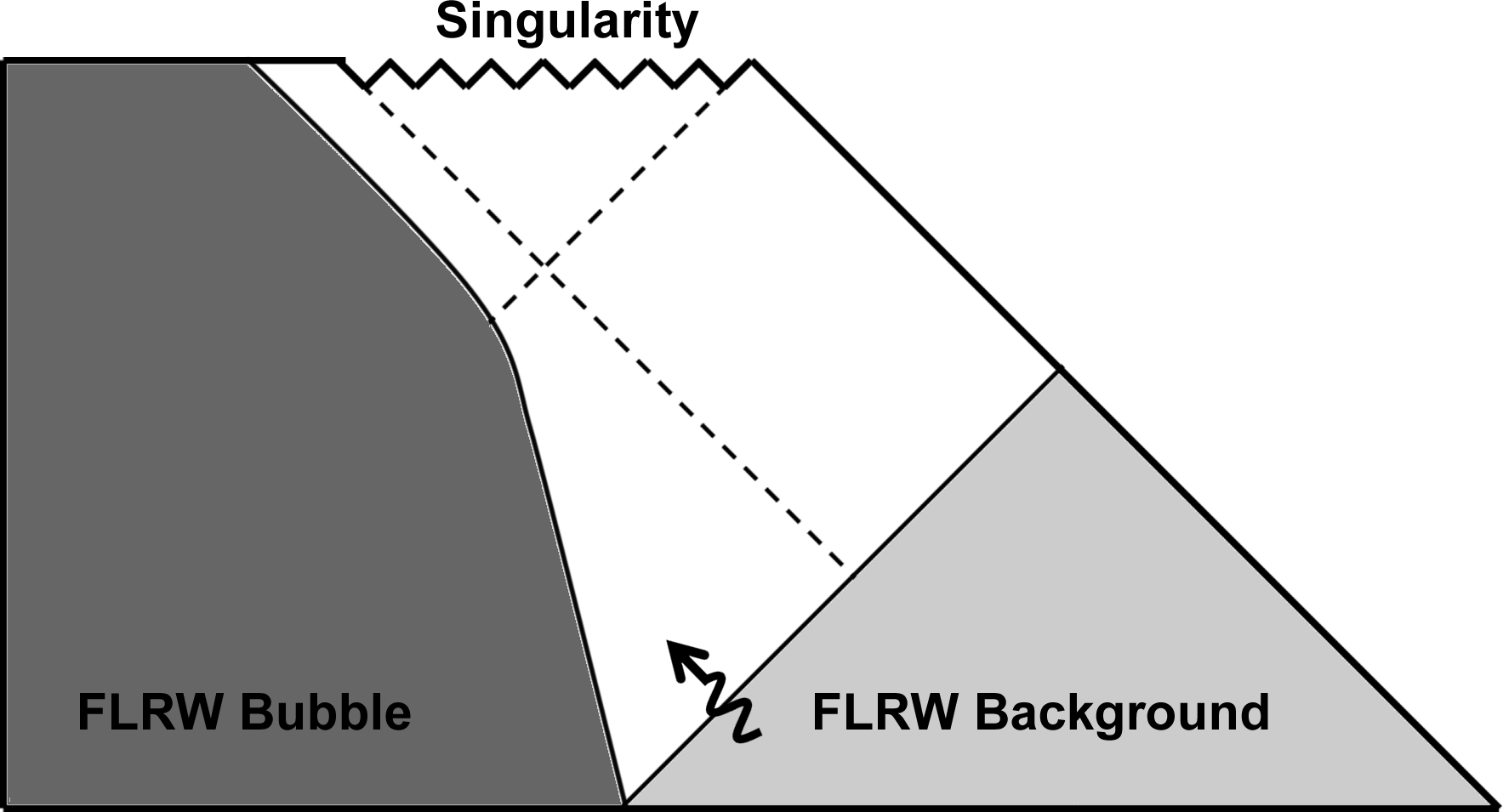}
\caption{One way to evade the constraints on bubble embeddings from the Raychaudhuri equation
is for the bubble to become disconnected from the background spacetime by the creation of a wormhole,
as shown in this Penrose diagram adapted from \cite{Garriga:2015fdk}. Radially ingoing null rays starting in the background
either see the bubble contracting or are casually disconnected from the bubble, while the bubble continues to expand
in a ``baby Universe''.
}
\label{fig:Wormhole_Penrose}
\end{figure}

It is possible to avoid the constraints we have derived here by embedding the cosmological bubble
in such a way that the bubble is casually disconnected from the background spacetime.
In particular, an expanding cosmological bubble tucked behind a wormhole
will be casually disconnected
from radially ingoing null rays from the background, as in Figure \ref{fig:Wormhole_Penrose}.
Solutions of this type have been described previously 
in \cite{Berezin:1982ur,Blau:1986cw,FARHI1990417} for a de-Sitter bubble and Schwarzschild
background, \cite{LagunaCastillo:1986je} for some limited FLRW spacetimes, 
\cite{Aurilia:1989sb,Aguirre:2005xs} for de-Sitter bubble and background spacetimes,
and \cite{Garriga:2015fdk,Deng:2016vzb} for false-vacuum and domain-wall bubbles embedded
in FLRW backgrounds. It is important to note, however, that the conclusions of \cite{Farhi:1986ty} still
apply, so that bubbles that are bigger than their apparent horizons still must begin in an initial singularity, even
if they are tucked behind a wormhole.

Finally, it would be useful to see how robust these arguments are to deviations from
homogeneous spherical symmetry in the bubble and the background (see \cite{Fischler:2007sz} for a discussion
of an inhomogeneous background).
We have also been assuming that the matter content of the background, bubble interior, and bubble wall
obeys the NEC; it is possible to consider cosmological models based on violations of the NEC,
either through quantum gravity effects \cite{Ford:1993bw}, NEC-violating matter fields (see
e.g.~\cite{ArkaniHamed:2003uz,Kobayashi:2010cm} for some common models), or non-minimal 
coupling \cite{Lee:2007dh,Lee:2010yd}.
We will leave a detailed study of bubbles with these effects for future work.

\section*{Acknowledgements}
The authors would like to thank T.~Ali and C.~Hellaby for useful discussions. We would also like to thank A.~Bhattacharyya for conversations on an earlier version of this work. SSH is supported by the Claude Leon Foundation. 


\bibliographystyle{utphysmodb}

\bibliography{refs}

\end{document}